\documentclass[aps,twocolumn,showpacs]{revtex4}
\usepackage{epsfig}
\usepackage{amsbsy}
\usepackage{amsmath}
\usepackage{amsfonts}
\usepackage{graphicx}

\begin{document}

\newcommand{\pf}{\noindent\textbf{Proof:~}}
\newcommand{\lemma}[1]{\noindent\textbf{Lemma {#1}\,:}}
\newcommand{\qed}{\hfill $\Box$}
\newcommand{\ba}{\begin{array}}
\newcommand{\ea}{\end{array}}
\newcommand{\me}{\ea\right)}
\newcommand{\mb}[2]{\left(\ba{#1}{#2}\ea\right)}
\def\sp#1{\langle #1 \rangle }
\newcommand{\be}{\begin{equation}}
\newcommand{\gcm}{\mbox{gcm}}
\newcommand{\ee}{\end{equation}}
\newcommand{\el}[1]{\label{#1}\end{equation}}
\newcommand{\erl}[1]{\label{#1}\end{eqnarray}}
\newcommand{\br}{\begin{eqnarray}}
\newcommand{\er}{\end{eqnarray}}
\newcommand{\for}{\qquad \mbox{ for}\quad}
\newcommand{\aand}{\qquad \mbox{ and}\quad}
\newcommand{\where}{\qquad \mbox{where}\quad}
\newcommand{\rf}[1] {(\ref{#1})}
\newcommand{\lb}[1] {\label{#1}}
\newcommand{\ie}{{\em i.e.~}}
\newcommand{\al}{\alpha}
\newcommand{\ot}{\otimes}
\newcommand{\Dc}{{\cal D}}
\newcommand{\Cc}{{\cal C}}
\newcommand{\Fc}{{\cal F}}
\newcommand{\Hc}{{\cal H}}
\newcommand{\Sc}{{\cal S}}
\newcommand{\Qc}{{\cal Q}}
\newcommand{\Tc}{{\cal T}}
\newcommand{\Pc}{{\cal P}}
\newcommand{\Uc}{{\cal U}}
\newcommand{\Zc}{{\cal Z}}
\def\tr{\mbox{tr\, }}
\def\kett#1{|#1 \rangle\! \rangle}
\def\ket#1{|#1\rangle}
\def\bra#1{\langle #1|}
\def\ker#1{|#1 )}
\def\brr#1{( #1|}

\title{Quantum gates on hybrid qudits}

\author{Jamil Daboul}
\affiliation{Department of Physics and Centre for Advanced Computing -- Algorithms and Cryptography,
Macquarie University, Sydney, New South Wales 2109, Australia}
\affiliation{Department of Physics, Ben-Gurion University, P.O.Box 653, Beer-Sheva 84105, Israel}
\author{Xiaoguang Wang}
\affiliation{Department of Physics and Centre for Advanced Computing -- Algorithms and Cryptography,
Macquarie University, Sydney, New South Wales 2109, Australia}
\author{Barry C. Sanders}
\affiliation{Department of Physics and Centre for Advanced Computing -- Algorithms and Cryptography,
Macquarie University, Sydney, New South Wales 2109, Australia}
\date{\today}

\begin{abstract}
We introduce quantum hybrid gates that act on qudits of different dimensions.
In particular, we develop two representative two-qudit hybrid gates
(SUM and SWAP) and  many-qudit hybrid Toffoli and Fredkin gates.
We apply the hybrid SUM gate to generating entanglement, and find that operator entanglement of the SUM gate is equal to the entanglement generated by it for certain initial states.
We also show that the hybrid SUM gate acts as an automorphism on the
Pauli group for two qudits of different dimension under certain conditions.
Finally, we describe a physical realization of these hybrid gates for spin systems.
\end{abstract}

\maketitle

%-------------------------------- 111 -------------------------------
\section{Introduction}
%----------------------------------------------------------------------

Although quantum computation is treated as processing qubits (quantum
versions of binary digits, or bits), quantum computing can be
generalized by considering logical elements of qudits (quantum versions
of $d$-ary digits)~\cite{Bry01}. Qubit-based quantum computation is
adequate for considering fundamental issues such as complexity classes
or computability, but, from a practical perspective, encoding as qudits
may be more natural, or constitute a more efficient use of
resources~\cite{Bar02}. For example, coupled harmonic oscillators can
admit various qudit encodings that exploit the full
Hilbert space~\cite{Got98,Bar02}.

Two-qudit gates have been treated, but so far always for two qudits of
equal dimensions~\cite{Bry01}. Here we treat hybrid qudit gates, namely
gates that transform two (or more) qudits of possibly different
dimensions. This analysis is particularly useful if two or more qudits
of different physical systems (and different dimensions) are coupled
together (such as a $d=2$ level system and a large $d$ dimensional
qudit in an oscillator). We develop two- and multi-qudit hybrid gates,
discuss possible physical realizations, and prove that the hybrid SUM gate
acts on the Pauli group for two qudits as an automorphism only when certain conditions on the dimensions of the qudit Hilbert spaces are met.

A {\em qudit} is a general state in a  $d$-dimensional Hilbert space
$\Hc_d$, \ie
$\ket{\Psi} = \sum_{m=0}^{d-1} c_m \ket{m}$, which reduces to
$\ket{\psi}= c_0 \ket{0} + c_1 \ket{1} $ for the qubit case.
A {\em basis} for a general multi-qudit system is given by
\br
&& |m_1 \rangle\otimes |m_2 \rangle\otimes
\cdots\otimes |m_N \rangle \cr
&& \quad \in  {\cal H}_{d_1}\otimes {\cal
H}_{d_2}\otimes\cdots\otimes {\cal H}_{d_N} ~,
~~m_i\in\mathbb{Z}_{d_i}.
\erl{hybridm}
If two or more $d_i$'s differ,
we refer to the multi-qudit system \rf{hybridm} as {\em `hybrid system'}.

This generalization is illuminating because it differs subtly from
standard non-hybrid qudit models
(see e.g. Lemma 2 in Sec. 5 below).
Moreover, hybrid systems have a
wider range of applications.  For example, a qubit can serve as a
control state with any qudit as the target state, or vice versa.
Also qubits are often only ideals: many systems involve multiple levels
for each degree of freedom, and the qubit is encoded into these levels.
The theory for hybrid qudit systems can be useful for
different interacting physical systems, with a $d_1$-dimensional
qudit natural for one system and a $d_2$-dimensional qudit natural for
another.

This paper is organized as follows. In Sec.~II we first review the
qudit computational basis and one-qudit operators. Then we
construct two hybrid versions of the SUM gate (see
Eqs.~\rf{sum1} and~\rf{alber} below)~\cite{Got98,Bar02,Alb00}, 
a partial-SWAP gate and a hybrid
version of the Toffoli~\cite{Tof80,Deu95,Cor98,Pri99,Wan012} and
Fredkin gates~\cite{Fre82,Mil89,Cha95,Chu95,Chu96} that were
instrumental in introducing the field of reversible (classical)
computation. In Sec.~III we calculate the operator entanglement of
the SUM gate and the entanglement generated by the SUM gate. In
Sec. IV we describe a realization of the hybrid gates by spin
systems. In Sec. V we prove a lemma that shows the SUM gate yields
an automorphism of the Pauli group by conjugation, if and only if
the dimension of the control system is a multiple of that of the
target system. We conclude in Sec. VI.

%-----------------------------------------------------------------
\section{Hybrid quantum gates }
%-----------------------------------------------------------------

%-----------------------------------------------------------------
\subsection{Generalized Pauli Group}
%-----------------------------------------------------------------

A basis for operators on ${\cal H}_d$ is given by the following
`generalized Pauli operators'~\cite{Pat88,Bar02,Kni96,Got98}
\begin{equation}
X^j Z^k,\quad j,k\in \mathbb{Z}_d,  \label{eq:PauliOperators}
\end{equation}
where $X$ and $Z$ are defined by their action on the computational basis
\begin{eqnarray}
X|s\rangle &=&|s+1\ ({\rm mod}\ d)\rangle \,,
\label{eq:ActionOfZOnCompBasis} \\
Z|s\rangle &=&\exp (2\pi {\rm i}s/d)|s\rangle \,=\zeta_d^s|s\rangle \,,
\end{eqnarray}
where
\begin{equation}
\zeta_d \equiv \exp (i2\pi /d)~.  \label{zeta}
\end{equation}
In the following we shall write for simplicity $\zeta$ instead of $\zeta_d$,
if the dimension is easily understood from the context.

The unitary operators $X$ and $Z$ generate the {\it generalized Pauli group} $\Pc_d$. Note that $X$ and $Z$ do not
commute; they obey
\be
Z^j X^k=\zeta^{jk} X^k Z^j~,
\el{cr}
and $X^d = Z^d=I.$

%-----------------------------------------------------------------
\subsection{One-qudit gates }
%-----------------------------------------------------------------

Before we consider two-qudit gates, we review some of
the properties of the useful one-qudit `Fourier gate' $F$,
which transfers the qudit computational basis $\ket{s}$ to the dual state
\begin{equation}
\ker{s} \equiv  F |s\rangle := \frac 1 {\sqrt{d}} \sum_{k=0}^{d-1}
\zeta^{sk} \ket{k} \for s\in\mathbb{Z}_d
\end{equation}
such that $\bra{s'}s)=1/\sqrt{d}~\zeta^{ss'}$. These dual states
are related to the computational basis by a discrete Fourier
transformation, and distinguished by a rounded bra/ket notation.
As an example, if
the computational basis corresponds to Fock number states for the
harmonic oscillator, the dual basis
corresponds to Susskind-Glogower phase states~\cite{Phase}.
Similarly the SU(2) phase states are dual to angular momentum eigenstates~\cite{Phase1}.

The $F$ gate is a qudit version of the one-qubit Hadamard gate
$H$. However, and in contrast to $H$, the $F$ operator for $d\ge 3$ is not
Hermitian and its order is 4 instead of 2, as~\cite{Stephen}
\be
F^2|s \rangle=|-s\rangle, \quad F^4 =I.
\ee
Similarly, the unitary operator $X$ can be considered as the qudit version
of the NOT gate, and $Z$ is the qudit version of the phase gate for qubits.

\subsection{Two-qudit gates}
%-----------------------------------------------------------------
\subsubsection{Hybrid SUM gate}
%-----------------------------------------------------------------

Two representative quantum gates on qubits are the controlled-NOT (CNOT) and
SWAP gate.
A generalised CNOT gate for qudits~\cite{Got98,Bar02,San02}
has been called the
displacement gate, or SUM gate~\cite{San02}.
As a compromise, we refer to the hybrid version of this `controlled-SHIFT'
operator as the `SUM gate', but use the notation $\Dc$ to emphasize
its displacement nature.
To achieve unity in notation, we shall use caligraphic letters to denote two- and multi-qudit gates. In particular, we shall use $\Sc, \Tc$ and $\Fc$ to denote the SWAP, the hybrid Toffoli, and Fredkin gates, respectively.

We now define the hybrid version of the SUM or displacement gate ${\cal D}$ on
${\cal H}_{d_c} \otimes {\cal H}_{d_t}$ for arbitrary $d_c$ and $d_t$ (the subscript $c$ refers to ``control'' and $t$ to ``target'') by
\begin{equation}
{\cal D} := \sum_{n=0}^{d_c-1} P_n \otimes X^n ~\qquad \mbox{ for}\quad
d_c,d_t \in \mathbb{N},
\el{sum1}
where
\begin{equation}
P_n \equiv |n \rangle\langle n|,~ n\in\mathbb{Z}_{d_c},  \label{p}
\end{equation}
is a {\em primitive projection operator} on  a computational basis state of the
{\em control space} ${\cal H}_{d_c}$.

It is important to note the following subtle difference between hybrid and
non-hybrid qudit systems: although the states
$\ket{i}\otimes \ket{j}$ and $\ket{i+ d_c}\otimes \ket{j}$
are formally equivalent, the operators
$P_{i}\otimes X^i=\ket{i}\bra{i}\otimes X^i$
and
$P_{i+d_c}\otimes X^{i+d_c} =\ket{i+d_c}\bra{i+d_c}\otimes X^{i+d_c}
=P_i\otimes X^{i+d_c}$ are {\em not} equal in general, if $d_c \ne d_t$.
Hence, in order to obtain a unique definition, we insist that
the summation in \rf{sum1} is restricted to $ 0 \le n < d_c$.
This subtle difference has interesting consequences when we try to define
a SWAP gate for hybrid systems.

For $d_c > d_t$ we can combine together all the projection operators $P_n$,
which yield the same $X^s$, and obtain
\begin{equation}
{\cal D} =\sum_{s=0}^{d_t-1} \Pi_s \otimes X^s~,
    \qquad \mbox{ for}\quad d_c
> d_t ~,
\el{sum2}
where
\begin{equation}
\Pi_s=\sum_{n=s \bmod d_c}^{d_c-1} P_n,\; \for s\in\mathbb{Z}_{d_t}.
\end{equation}
For example, the SUM gate for $d_c=3$ and $d_t=2$ is given by
$$
{\cal D} = \sum_{s=0}^1 \Pi_s \otimes X^s = \Pi_0 \otimes I + \Pi_1
\otimes X ~,
$$
where $\Pi_0= P_0 + P_2$ and $\Pi_1= P_1 $.

We can extend expression \rf{sum2}, also
for $d_c \le d_t$, by defining
\be
\Dc :=\sum_{s=0}^{d_{\min}-1} \Pi_s \otimes X^s~,
\el{sum3}
where $d_{\min}:=\min(d_c,d_t)$.
Note that $\sum_{s=0}^{d_{\min}-1} \Pi_s= I_{d_c\times d_c}$~.

We introduce another interesting hybrid gate:
\br
&&
\Dc'_{12}|m\rangle\otimes|n\rangle :=|m\rangle\otimes |m-n\rangle, \cr
&& \for m\in\mathbb{Z}_{d_c} \aand n\in\mathbb{Z}_{d_t} ~.
\erl{alber}
This operator is unitary and Hermitian, as $(\Dc'_{12})^2=I$.
It is related to the SUM gate by

$$\Dc'_{12}=\Dc_{12} ~(I\otimes F^2)~.$$
For $d_c=d_t$ our hybrid $\Dc'_{12}$ reduces to the generalized CNOT gate given by Alber et al.~\cite{Alb00}.

%-----------------------------------------------------------------
\subsubsection{The SWAP gate}
%-----------------------------------------------------------------

The SWAP operation on $\Hc_d\times \Hc_d$ systems,
\ie for $d_c=d_t=d$ systems, is defined by
\be
\Sc \ket{i}\otimes\ket{j}= \ket{j}\otimes\ket{i}~,
\for i,j\in\mathbb{Z}_d~,
\ee
hence, $\Sc = \sum_{i,j=0}^{d-1}
\ket{j}\bra{i} \otimes \ket{i}\bra{j}$. Clearly, the definition cannot be used
for hybrid systems. Instead, for $d_c \ne d_t$ (and also for $d_c = d_t$)
we define partial-SWAP operators by
\be
\Sc_P  \ket{i}\otimes\ket{j}= \left\{\ba{ll}
\ket{j}\otimes\ket{i}~, &\for i,j\in\mathbb{Z}_{d_P}, \\
\ket{i}\otimes\ket{j}~, &~\qquad \mbox{otherwise} \ea \right.
\el{sw3}
where $d_P \le d_{\min}= \min (d_c,d_t)$.
Obviously, $\Sc_P$ in \rf{sw3} is unitary and Hermitian, as $\Sc_P^2=I$.
This partial SWAP gate only acts as a SWAP operation
on a subspace of the original Hilbert space.

\subsubsection{Relation between SWAP and SUM operators}

It is easy to check that $\Sc$ can be written in terms
of three SUM gates as follows
\be
\Sc = (F^2\otimes I){\cal D}_{12}{\cal D}_{21}^{-1}{\cal D}_{12}.
\el{sw1}
Another possibility is to use expressions \rf{sw1}
formally to define a swap-like gate for hybrid system.
However, contrary to what one might expect, this operator does
not yield a swap operation, even for $0\le i,j \le d_{\min}$.

We illustrate this claim by a simple example, where $d_1=3$ and $d_2=2$.
By applying expression \rf{sw1} to the state
$\ket{0}\otimes \ket{1}$. We obtain successively
\begin{eqnarray}
&\ket{0}\otimes \ket{1} \longrightarrow
\ket{0}\otimes \ket{1} \longrightarrow
\ket{2}\otimes \ket{1} \nonumber\\
& \longrightarrow
\ket{2}\otimes \ket{1}
 \longrightarrow
%\ket{-2}\otimes \ket{1} =
\ket{1}\otimes \ket{1} \ne
\ket{1}\otimes \ket{0}
\end{eqnarray}

Recently,
Fujii constructed a swap gate, as follows~\cite{Fuj02}
\begin{equation}
\Sc={\cal D}_{12}(F^2\otimes I) {\cal D}_{21}(F^2\otimes I)
{\cal D}_{12}(I\otimes F^2),
\el{sw2}
expressed in our notations. Note that both constructions of SWAP gates actually require three SUM gates
and three local $F^2$ gates. This is because
\be
\Dc_{21}^{-1}=(I \otimes F^2) \Dc_{21} (I\otimes F^2),
\ee
so that our SWAP gate \rf{sw1} can be written as
\be
\Sc= (F^2\otimes I){\cal D}_{12}(I\otimes F^2) \Dc_{21} (I\otimes F^2)
{\cal D}_{12}.
\ee
We also note that the SWAP gate on continuous variables can be
constructed by three generalized controlled-NOT gates on continuous variables~\cite{Wan01}.

\subsection{Higher order quantum hybrid gates}
%-----------------------------------------------------------------

Representative higher-order three-qubit gates include the quantum
versions of the Toffoli gate~\cite{Tof80,Deu95,Cor98,Pri99,Wan012} and
of the Fredkin gate~\cite{Fre82,Mil89,Cha95,Chu95,Chu96};
these three-bit gates are important primitives for
logically reversible classical computation,
for which universal reversible two-bit gates do not exist.
The Toffoli gate is effectively a controlled-controlled-NOT (C$^2$NOT),
and the Fredkin gate is another universal three-bit gate.

As a controlled-controlled-NOT,
the quantum Toffoli gate has two qubits as control and one qubit as target, and the target qubit flips if and only the two
 control qubits are in the state $|1\rangle\otimes|1\rangle$. The Fredkin gate has one qubit as control and two qubits as
 target, and the states of two target qubits swap if and only if the control qubit is in the state $|1\rangle$.
Here we give the hybrid version of these two higher-order gates.

\subsubsection{The hybrid Toffoli gate}

A general controlled unitary gate acting on Hilbert spaces ${\cal H}%
_{d_c} \otimes {\cal H}_{d_t}$ can be written as
\begin{equation}
\Cc_U = \sum_{s=0}^{d_c-1} P_s \otimes U_s = \sum_{s=0}^{d_c-1}
\ket{s}\bra{s} \otimes U_s~,  \label{cu}
\end{equation}
where $U_s$ are arbitrary unitary operators on the {\em target space} ${\cal %
H}_{d_t}$.

Note that $\{U_s\}$ may be unitary operators on single or multiple qudits, and may include the case of qudit-controlled
operators on other qudits. The latter case allows unitary operators on qudits that can be jointly controlled by two or
more qudits.
An example is provided by the following
`natural' generalization of the Toffoli gate
~\cite{Tof80,Deu95,Cor98,Pri99,Wan012}
\begin{equation}
{\cal T}:=\sum_{s=0}^{d_c-1}P_s\otimes {\cal D}^s~,  \label{t}
\end{equation}
where the $U_s$ in (\ref{cu}) are replaced by ${\cal D}^s$, which are powers
of the generalized displacement operator (\ref{sum1}). The hybrid Toffoli-type
gate is thus a {\em `triple gate'}
\begin{equation}
{\cal T}=\sum_{r=0}^{d_c-1}\sum_{s=0}^{d_c^{\prime }-1}P_r\otimes P_s\otimes
X^{rs}=\sum_{m=0}^{d_t-1}\Pi_m\otimes X^m~,  \label{tp}
\end{equation}
where $\Pi _m$ are compound projection operators, given by
\begin{equation}
\Pi _m=\sum_{r=0}^{d_c-1}\sum_{s=0}^{d_c^{\prime }-1}\delta
_{m,rs}~P_r\otimes P_s~,\qquad m\in\mathbb{Z}_{d_t}~.  \label{pi}
\end{equation}
where the products $rs$ of the delta in (\ref{pi}) are defined modulo $d_t$. Hence, the order of the Toffoli gate is equal
to $d_t$.

%-----------------------------------------------------------------
\subsubsection{The hybrid Fredkin gate}
%-----------------------------------------------------------------

Another type of multi-qudit gate is the quantum Fredkin gate~\cite{Fre82}--\cite{Chu96}. We define the
{\em hybrid Fredkin gate} on
$ \Hc_{d_c}\ot \Hc_{d_1}\ot \Hc_{d_2}$  by
\be
\Fc := \sum_{m=0}^{d_c-1}P_m\otimes \Sc_P^m = \Pi_+ \ot I + \Pi_- \ot \Sc_P
\ee
where $\Pi_\pm$ are the following projection operators
\be
\Pi_+ :=\sum_{{\rm even}\ m}P_m  \aand  \Pi_-:=\sum_{{\rm odd}\ m}P_m
\ee
where we have used the property $\Sc_P^2=I$.
%and employ the notation that $m|2$ means `$m$ divides 2' and hence
%is even; the notation $m\!\not\;\mid 2$ means that `$m$ is odd'.

The hybrid Fredkin gate
executes a swap for purely odd state $\ket{\psi_-}$, \ie for
$ \Pi_-\ket{\psi_-}=\ket{\psi_-}$, and does nothing for the even states.
However, for mixed odd and even states, one obtains a mixed result.
For instance, if we choose a input state as $(|0\rangle+|1\rangle)\otimes|\alpha\rangle\otimes|\beta\rangle$, the output
state after the gate is $|0\rangle\otimes|\alpha\rangle\otimes|\beta\rangle+|1\rangle\otimes|\beta\rangle\otimes|\alpha\rangle$,
which is in general an entangled state.

%-----------------------------------------------------------------
\section{Entanglement produced by  quantum gates}
%-----------------------------------------------------------------

Hybrid two- and multi-qudit gates can enhance entanglement, i.e. the entanglement of the output state can be greater than
that of the input state.
In this case we regard the hybrid gates as entangling gates. Different methods exist for characterizing the enhancement of
entanglement. In this section, we discuss entanglement enhancement by the hybrid SUM gate.

%-----------------------------------------------------------------
\subsection{Entanglement measures for states and operators}
%-----------------------------------------------------------------
There are various measures of entanglement for a {normalized
state} $\ket{\psi} \in \Hc_{d_1}\ot \Hc_{d_2}$. Here, we shall use
the von Neumann entropy
\be
E(\ket{\psi})= - \sum_{n=0}^{N_S-1} p_n \log p_n~.
\el{sch}
where $\{p_n\}$ is defined in terms of the Schmidt decomposition
of $\ket{\psi}$:
\be
\ket{\psi}
=\sum_{n=0}^{N_S-1} \sqrt{p_n} |\phi_n\rangle\otimes |\chi_n\rangle, \quad p_n > 0~\forall n,
\el{sch0}
and $\log$ is always taken to be base 2.
Definition~\rf{sch} was adapted~\cite{Zan01,Nie02}
to define {\em operator entanglement}, as follows.
Let $\Qc$ be an operator acting on a hybrid space $\Hc_{d_1}\ot \Hc_{d_2}$,
with the following Schmidt decomposition~\cite{Nie02}
\be
\Qc=\sum_{n=0}^{N_S-1} s_n A_n\otimes B_n,
\el{schsch}
with $s_n > 0~\forall n$, and the two operators $A_n$ and $B_n$ are orthonormal with respect to the Hilbert-Schmidt scalar
product defined by $\langle A,B \rangle := {\tr (A^\dagger B)}$ for $A$ and $B$ two arbitrary operators.
In particular, $||A|| := \sqrt{\tr (A^\dagger A)}$ is the Hilbert-Schmidt norm of
the operator $A$, and $\hat {A}:= A/||A||$ if $||A||\neq 0$.

Since linear operators over a finite-dimensional vector space
${\cal H}_d$ can be regarded as $d^2$-dimensional vectors,
we may think of $\widehat {\Qc} \equiv \Qc/||\Qc||$ as a
normalized state, which we denote by $\kett{\widehat {\Qc}}$,
so that \rf{schsch} becomes
\be
\kett{\widehat {\Qc}}=\sum_{n=0}^{N_S-1} \sqrt{p_n}~
\kett{A_n}\otimes \kett{B_n},
\el{sch2}
where $\sqrt{p_n}=s_n/||\Qc||$~.
In particular, if $\Qc$ is unitary, then
$||\Qc||=\sqrt{d_cd_t}$. The operator entanglement~\cite{Nie02}
\begin{equation}
E_{\text{op}}(\Qc)
= - \sum_n \frac{s_n^2}{d_cd_t}\log \left( \frac{s_n^2}{d_cd_t}
\right)~.\label{sch22}
\end{equation}

%-----------------------------------------------------------------
\subsection{Operator entanglement of the SUM gate}
%-----------------------------------------------------------------

In Sec.~II, we essentially obtained in Eq.~\rf{sum3}
the Schmidt decomposition of the operator $\Dc$ because the projection operators $\Pi_s$ and the unitary operators
$X^s$ are mutually orthogonal, \ie
\br
\langle \Pi_r,\Pi_s \rangle &=& ||\Pi_s||^2 \delta_{r,s}, ~r,s\in\mathbb{Z}_{d_{\min}}\lb{prs} \\
\langle X^r, X^s \rangle &=& ||X^s||^2 \delta_{r,s}= d_t \delta_{r,s}, ~r,s \in\mathbb{Z}_{d_t},
\erl{xrs}
where we used $||X^s||^2 = \tr (X^{s\dagger}X^s ) =\tr I=d_t$, because
$X^s$ is unitary.
Hence, by dividing the operators $\Pi_s$ and $X^s$ in \rf{sum3} by their
norms, we immediately obtain the following Schmidt decomposition of $\Dc$:
\be
\Dc :=\sum_{s=0}^{d_{\min}-1} (||\Pi_s||\, \sqrt{d_t})~\widehat{\Pi}_s \otimes
\widehat{X}^s~,
\el{sum4}
where for $d_c=Kd_t+r$ we have
\be
||\Pi_s||
% &=& \sqrt{\tr (\Pi_s^\dagger \Pi_s)}=
%\sqrt{\tr (\Pi_s)}  \for 0\le s\le d_{\min}-1\cr
= \left\{\ba{ll} \sqrt{K+1} &(0\le s\le r-1), \\
\sqrt{K} &(r\le s\le d_t-1). \ea \right.
\el{npis}

>From Eqs.~\rf{sum4} and \rf{npis} expression \rf{sch2} yields immediately
\be
E_{\text{op}}(\Dc)= e_\Dc(d_c,d_t)~,
\ee
where (for  $d_c=Kd_t+r$)
\be
e_\Dc(d_c,d_t)=-r\frac{K+1}{d_c}\log \frac{K+1}{d_c} -(d_t-r)\frac K{d_c}\log
\frac K{d_c}~.
\el{ent}
Note that for $d_c<d_t$ the general expression \rf{ent} reduces
simply to
\be
e_\Dc(d_c,d_t) = \log d_c~, \for  d_c<d_t,
\el{sch222}
by substituting $K=0$ and $r=d_c$.

%-----------------------------------------------------------------
\subsection{Entanglement produced by the SUM gate}
%-----------------------------------------------------------------

We prove the following lemma:\\

\lemma {1} {\em  The entanglement generated by the hybrid SUM gate $\Dc$
on the following three initial product states (one without and
two with ancillas)
\br
\ket{\Psi_1} \equiv
\ket{\gamma} \ot \ket{t}
&=& \left(\frac 1 {\sqrt{d_c}} \sum_{m=0}^{d_c-1}|m\rangle \right) \ot \ket{t},
 \lb{ga} \\
\ket{\Psi_2} \equiv
\ket{\al} \ot \ket{t} &=& \left(\frac{1}{\sqrt{d_c}}
\sum_{m=0}^{d_c-1}|m\rangle\ot \ket{m}\right) \ot \ket{t},  \lb{al} \\
\ket{\Psi_3} \equiv
\ket{\al} \ot |\beta\rangle &=& \ket{\al} \ot \left(\frac{1}{\sqrt{d_t}}
\sum_{n=0}^{d_t-1}|n\rangle\otimes|n\rangle \right),
\erl{ins}
where $\ket{t}$ is any of the computational states of the target space,
are equal to the operator entanglement \rf{ent} of $\Dc$} \ie
\be
E(\Dc\ket{\Psi_k}) = E_{\text{op}}(\Dc)= e_\Dc(d_c,d_t)~.
\el{e}

\pf The three initial states have zero entanglement, since they were
chosen to be  {\em product} states.
Therefore, the increase of entanglement due
to $\Dc$ is equal to  $E(\Dc\ket{\Psi_k})$.

We shall now apply $\Dc$ to \rf{ga} :
\be
|\Psi^f_1 \rangle \equiv {\cal D}~|\gamma\rangle \ot \ket{t}
=\left(\frac 1{\sqrt{d_c}}\sum_{s=0}^{d_{\min}-1}\Pi_s|m\rangle \right) \otimes
X^s |t\rangle.
\el{f2}
Let $d_c=Kd_t+r $~ (Note that $K=0$ and $r=d_c$ if $d_c<d_t$).
Hence,
\br
&&\sum_{m=0}^{d_{\min}-1} \Pi_s \ket{m} =\cr
&&\left\{
\begin{array}{l}\ket{s}+
|s+d_t\rangle +\ldots +|s+Kd_t\rangle =\sqrt{K+1}|\psi _s\rangle \\
\quad \qquad \mbox{ for}\quad 0\le s\le r-1, \\
|s\rangle +\ldots +|s+(K-1)d_t\rangle =\sqrt{K}|\psi _s\rangle \\
\quad \qquad \mbox{ for}\quad r\le s\le d_t-1~,
\end{array}
\right.
\erl{states}
where the $|\psi_s\rangle, ~s\in\mathbb{Z}_{d_\text{min}}$
are orthonormal states
which, for $d_t<d_c$, span a $d_t$--dimensional {\em subspace} of ${\cal H}_{d_c}$. By
substituting \rf{states} into \rf{f2}, we obtain
the following Schmidt decomposition of the final state
\be
\ket{\Psi^f_1}={\cal D}~|\gamma\rangle \otimes \ket{t} =
\sum_{s=0}^{d_c-1} \sqrt{p_s} |\psi_s\rangle \ot \ket{t+s}
\el{f3}
where
\be
p_s =\left\{\ba{ll} (K+1)/d_c &\for 0\le s\le r-1, \\
K/d_c &\for r\le s\le d_t-1, \ea \right.
\ee
By substituting the above equation into \rf{sch0} we obtain exactly the same
expression \rf{ent}. Similarly, we can prove that the entanglement of
$E(\Dc|\alpha\rangle\otimes|t\rangle)$ is also given by \rf{ent}.

Finally, since the states 
$\{X^s \ket{\beta} \}$
are orthonormal for different $s$,
we get essentially the same Schmidt decomposition for $\Dc\ket{\al}\ket{\beta}$
as in \rf{f3}, and hence the same final entanglement.
This result also follows from lemma 5 of Ref.~\cite{Nie02}. \qed \\

%\vspace{2mm}
%
The entanglement function~(\ref{ent}) is plotted in Fig.~1.
As the generated entanglement equals the operator entanglement
according to Eq.~(\ref{e}), Fig.~1 presents $E$ as the ordinate axis.
We observe in Fig.~1 that the entanglement approaches log$_2d_t$
as $d_c$ becomes large. We can see this asymptotic result in Eq.~(\ref{ent})
by noting that
$$
\frac{K+1}{d_c}=\frac{d_c+d_t-r}{d_cd_t}\rightarrow\frac{1}{d_t}
$$
so the entanglement asymptotically approaches log$d_t$ as observed
in Fig.~1.

\begin{figure}
\begin{center}
\epsfig{width=10cm,file=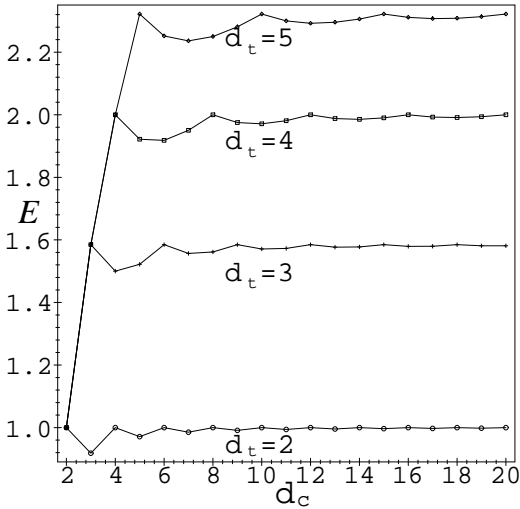}
\caption[]{The operator entanglement of the hybrid SUM gate for different
$d_c$ and $d_t$.}
\end{center}
\end{figure}

\section{Physical realization of hybrid gates}

One can encode a qudit in physical systems such as spin systems and harmonic oscillators~\cite{Bar02}. The Hilbert space
associated with a spin--$j$ system is spanned by the basis $\{|j,m\rangle
;m=-j,\cdots,j\}$ and the su(2) algebra is generated by $\{J_x,J_y,J_z\}$,
with $[J_x,J_y]=iJ_z$ etc.\
and $(J_x^2+J_y^2+J_z^2)|j,m\rangle=j(j+1)|j,m\rangle$.
It is natural to define a number operator $N$ and number states as follows
\br
N &:=&J_z+j I, \\
|n\rangle _j &:=&|n-j\rangle \quad (n=0,\cdots,2j).
\end{eqnarray}
Then we have $N \ket{n}_j=n \ket{n}_j.$ In the spin system the operators
$X$ and $Z$ are realized as
\br
X &=&\sum_{n=0}^{2j}|n+1\rangle _{j\,j}\langle n|, \\
Z &=&\exp[{i2\pi N/(2j+1)}].
\end{eqnarray}

\subsection{Controlled-phase and SUM gates}

We consider interaction between spin--$j_1$ and spin--$j_2$ systems, via the
Hamiltonian $H=-gJ_{cz}J_{tz}$.
Up to local unitary operators the evolution operator
$\exp (itgJ_{cz}J_{tz})$
is equivalent to
$U(t)=\exp (itgN_cN_t)$.
By choosing $tg=2\pi /(2j_t+1)=2\pi /d_t$, we obtain
the unitary operator
\be
V= \exp\Big[{i\frac{2\pi }{d_t}N_cN_t}\Big]=\zeta_{d_t}^{N_cN_t},
\end{equation}
which is just the controlled-phase gate~\cite{Got98}. On the other hand, we know that the SUM gate can be obtained from the controlled-phase gate as follows~\cite{WangBarry02}
\begin{equation}
{\cal D}=(I\otimes F^\dagger)\zeta_{d_t}^{N_cN_t}(I\otimes F).
\end{equation}
Therefore, with the aid of $F$ gate we realized the hybrid SUM gate.

\subsection{Toffoli gate}
Now let us see how to physically create a hybrid Toffoli gate. Refs.~\cite{Wan012,WanSimu} shows that the interaction
Hamiltonian $N_1 N_2 N_3$ ($N_i$ correspond to spin-$j_i$ and one $j_i$  is equal to 1/2)
arises in ion-trap systems when coupling these operators $N_i$ to a common continuous variable.
The dimension of a spin-$j_i$ system is given by $d_i =2j_i+1$.
Therefore, we have the three-body controlled-phase gate
\begin{equation}
W(\theta)=e^{i\theta N_1N_2N_3}
\end{equation}

By choosing, say, $\theta=2\pi/d_3$, we make $\Hc_{d_3}$ the target space while
$\Hc_{d_1}\otimes\Hc_{d_2}$ becomes the control space.
Then, by appending the appropriate $F$ gate on the target system, we can realize the Toffoli gate acting on the systems
$\Hc_{d_1}\otimes \Hc_{d_2}\otimes \Hc_{d_3}$.

\subsection{Fredkin gate}

As a final remark we point out that we can construct a control-SWAP gate that acts on
$\Hc_{d}\otimes \Hc_{\infty}\otimes \Hc_{\infty}$ as a generalization of the controlled-SWAP gate acting on
 $\Hc_{2}\otimes \Hc_{\infty}\otimes \Hc_{\infty}$ system~\cite{Wan01}.

The SWAP gate between two bosonic modes $a_1$ and $a_2$
is given by~\cite{Wan01}
\begin{equation}
\Sc_{12}=e^{i\pi a_2^{\dagger }a_2}e^{\frac \pi 2(a_1^{\dagger}a_2-a_2^{\dagger }a_1)}.
\label{eq:swapp}
\end{equation}
In an ion-trap system we can couple the spin-$j$ system to two bosonic modes $a_i$ $(i=1,2)$  as \cite{Ion1,Ion2}
\be
H_i=\chi N a_i^\dagger a_i
\label{eq:hi}
\ee
Since operators $H_i$ commute with each other, we can simulate the following Hamiltonian
\begin{equation}
H=H_1-H_2=\chi N (a_1^{\dagger }a_1-a_2^{\dagger }a_2)=2\chi N J_z ,
\end{equation}
where $J_z=\frac 12(a_1^{\dagger }a_1-a_2^{\dagger }a_2)$.
The operators
$J_z$ and $J_+=a_1^{\dagger }a_2=J_-^\dagger$ form the
su(2) Lie algebra. The evolution operator of the Hamiltonian $H$ at
time $t=-\pi /2\chi $ is given by
\be
U=U(-\pi/2\chi)=e^{i\pi J_zN}.
\end{equation}
The evolution operator $U$ can be transformed to $U^{\prime }\,$as
\be
U^{\prime }=e^{i\frac \pi 2J_x}Ue^{-i\frac \pi 2J_x}=e^{i\pi J_yN}= e^{\frac \pi 2 N(a_1^{\dagger}a_2-a_2^{\dagger }a_1)}
\label{eq:uprime}
\ee
where $J_x=(J_{+}+J_{-})/2$ and $J_y=\left( J_{+}-J_{-}\right) /(2i)$.
>From Eqs.~(\ref{eq:swapp}), (\ref{eq:hi}), and (\ref{eq:uprime}), we construct  the controlled-SWAP gate (hybrid Fredkin gate) as
\begin{eqnarray}
\Fc&=&e^{i\pi a_2^{\dagger }a_2N}e^{i%
\frac \pi 4(a_1^{\dagger }a_2+a_2^{\dagger }a_1)}  \nonumber \\
&&\times e^{i\frac \pi 2a_1^{\dagger }a_1N}e^{-i\frac \pi 2%
a_2^{\dagger }a_2N}e^{-i\frac \pi 4(a_1^{\dagger }a_2+a_2^{\dagger
}a_1)} \nonumber\\
&=&\Sc^{N}.
\end{eqnarray}
Therefore we have provided a controlled-SWAP gate on $\Hc_d\otimes\Hc_\infty\otimes\Hc_\infty$ systems in terms of
five two-body operators.

%-----------------------------------------------------------------
\section{Conjugation by the SUM gate}
%-----------------------------------------------------------------

A conjugation by the SUM gate $\Dc$ is described by the following lemma:\\

\lemma {2} {\em  The hybrid SUM gate $\Dc$ yields, by conjugation,
an automorphism of the Pauli group $\Pc_{d_c}\ot \Pc_{d_t}$, iff
$d_c/d_t$ is an integer $K$.
More explicitly, }
\br
{\cal D}(X\otimes I){\cal D}^{\dagger } &=& X\otimes X \lb{cr1} \\
{\cal D}(I\otimes X){\cal D}^{\dagger } &=& I\otimes X \lb{cr2} \\
{\cal D}(Z\otimes I){\cal D}^{\dagger } &=& Z\otimes X \lb{cr3} \\
{\cal D}(I\otimes Z){\cal D}^{\dagger } &=&
\left(\sum_{s=0}^{d_c-1} \zeta_{d_c}^{-sd_c/d_t}P_s\right) \otimes Z
\lb{cr4} \\
&=& Z^{-K} \ot Z \for \frac {d_c}{d_t} =K~.
\erl{cr5}
\vspace{3mm}
\pf
By noting that
\begin{equation}
P_r X P_s= P_r \ket{s+1}\bra{s}=
\ket{s+1}\bra{s}~ \delta_{r,s+1},
\end{equation}
we obtain
\begin{align}
{\cal D}(X\otimes X^k){\cal D}^{\dagger }
&= \sum_{s=0}^{d_c-1} P_r X P_s\otimes X^{r+k-s}
    \nonumber   \\
&=X\otimes X^{k+1}.
\end{align}
This proves both \rf{cr1} and \rf{cr2} simultaneously.
By noting that $Z^j=\sum_{s=0}^{d-1}\zeta_{d}^{sj} P_s$, we get
\begin{equation}
{\cal D}(Z\otimes I){\cal D}^{\dagger }
    =\sum_{r,s,t=0}^{d_c-1}\zeta_{d_c}^s P_rP_sP_t
    \otimes X^{r-t} = Z\otimes I.
\label{czcx}
\end{equation}
Finally, by using the commutation relation \rf{cr} and
$\zeta_{d_t}=(\zeta_{d_c})^{d_c/d_t}$, we obtain
\br
{\cal D}(I\otimes Z){\cal D}^\dagger
&=& \sum_{s=0}^{d_c-1} P_s\otimes X^s Z X^{-s}
= \sum_{s=0}^{d_c-1} P_s\otimes  \zeta_{d_t}^{-s} Z\cr
&=& \sum_{s=0}^{d_c-1}\zeta_{d_c}^{-s d_c/d_t} P_s \otimes Z \lb{zz} \\
&=& Z^{-K}\otimes Z~, \for  d_c= K d_t~.
\er
\qed

Note that even if $d_c/d_t =K\ge 2 $ is an integer, then $\Dc_{12}$ but not
$\Dc_{21}$ will belong to the Clifford algebra of the hybrid Pauli group.

\section{Summary}

We considered quantum hybrid gates which act
on tensor products of qudits of different dimensions. In particular,
we constructed two-body hybrid SUM and partial-SWAP gates,
and also many-body hybrid Toffoli and  Fredkin gates.
We have calculated the entanglement generated by the SUM gate.
We describe a physical realization of these hybrid gates for spin systems.
We also proved two lemmas, one related to entanglement generation with and
without ancillas, and the other involving conjugation by the SUM gate.\\

\acknowledgments

Jamil Daboul thanks Macquarie University for its hospitality. We appreciate
valuable discussions with Stephen Bartlett and Dominic Berry. This project
has been supported by an Australian Research Council Large Grant
and by a Macquarie University Research Grant.

%%%%%%%%%%%%%%%%%%%%%%%%%%%%%%%%%%%%%%%%%%%%%%%%%%%%

\end{document}